\pdfoutput=1 
\documentclass{JINST}

\usepackage{units}
\usepackage{amsmath}
\usepackage{amssymb}

\usepackage{upgreek}
\usepackage{cases}
\usepackage{lineno}

\usepackage{subcaption}

\title{Precision measurement of the carrier drift velocities in <100> silicon}

\author{C. Scharf\thanks{Corresponding author.~}\,\, and R. Klanner

\\
\llap{}University of Hamburg,\\
  Luruper Chaussee 149, 22761 Hamburg, Germany\\
E-mail: \email{Christian.Scharf@desy.de}}

 \abstract{Measurements of the drift velocities of electrons and holes as functions of electric field strength and temperature in high-purity n- and p-type silicon with <100> crystal orientation are presented.
 The measurements cover electric field strengths between 2.4 and 50~kV/cm and temperatures between 233 and 333~K.
 Two methods have been used for extracting the drift velocities from current transient measurements:
 A time-of-flight (tof) method and fits of simulated transients to the measured transients, with the parameters describing the field and temperature dependence of the electron and hole mobilities as free parameters.
 A new mobility parametrization, which also provides a better description of existing data than previous ones, allowed an extension of the classical tof method to the situation of non-uniform field strengths.
 For the fit method, the use of the convolution theorem of Fourier transforms enabled us to precisely determine the electronics transfer function of the complete set-up, including the sensor properties.
 The agreement between the tof and the fit method is about 1~\%, which corresponds to a time-of-flight uncertainty of 30~ps for a pad diode of $\unit[200]{\upmu m}$ thickness at the highest voltages.
 Combining our results with published data of low-field mobilities, we derive parameterizations of the drift velocities in high-ohmic <100> silicon for electrons and holes for field strengths between 0 and 50~kV/cm and temperatures between 233 and 333~K.}


\begin{document}

\section{Introduction}\label{sec:intr}

 The understanding and accurate simulation of silicon sensors requires the precise knowledge of basic material parameters, like the drift velocity of the charge carriers as a function of the electric field strength.
 At present, most silicon sensors used for tracking in high energy physics or for imaging in X-ray science use silicon with <100> lattice orientation, for which the field dependence of the drift velocity is only poorly known~\cite{jac1977,bec2011}, and parameterizations based on measurements exist only up to 20~kV/cm.
 Therefore, we performed a precision measurement of the drift velocities in <100> silicon for electric field strengths between 2.4 and 50\,kV/cm and temperatures between 233 and 333\,K, which covers  the operating conditions at high luminosity colliders.
 This paper summarizes the main results of our study.
 More details are given in Refs.~\cite{KS2015,scharf2014thesis,transferf}.

\section{Mobility Parametrization}\label{sec:mobility}

 As we have not been able to describe the measurements, which are reported in the next section, with the commonly used Caughey-Thomas~(CT) parametrization~\cite{caughey1967carrier}, we investigated the time-of-flight (tof) data published by Canali et al.~\cite{canali1971drift} for both the <111> and the <100> crystal directions.
 We found that a new parametrization of the carrier mobility $\mu(E)$ as a function of the electric field strength $E$ with a threshold field $E_0$ improves the description of the data:
\begin{eqnarray}
   1/\mu^{KS}(E) & =
   \begin{cases}
   \nicefrac{1}{\mu_{0}^{KS}} & E<E_{0}\\
   \nicefrac{1}{\mu_{0}^{KS}}+\nicefrac{1}{v_{sat}^{KS}}\cdot(E-E_{0}) & E\geq E_{0}\,,
   \end{cases}
  \label{eq:newparlin}
 \end{eqnarray}
 with the low-field mobility $\mu_0$ and the saturation velocity $v_{sat}$.
 The root-mean-square deviation between fit and data is reduced by about a factor two using  Eq.~\ref{eq:newparlin}.

 The minimum field strength of our measurements is 2.4~kV/cm, which is above the threshold field strength $E_0 \approx $ 1.8~kV/cm for electrons at 300~K, as determined from the data of Ref.~\cite{canali1971drift}.
 Thus the parametrization of the mobility used for electrons is identical to the  Trofimenkoff~\cite{trofimenkoff1965field} (Tr) parametrization:
 \begin{eqnarray}
   1/\mu^{Tr}(E) & = & \nicefrac{1}{\mu_{0}^{Tr}}  + \nicefrac{E}{v_{sat}^{Tr}}.
  \label{eq:munewe}
 \end{eqnarray}
 For holes the description of the data is significantly improved when a quadratic term is introduced:
 \begin{eqnarray}
  1/\mu_{h}(E) & =
  \begin{cases}
  1/\mu_{0}^{h} & E<E_{0}\\
  1/\mu_{0}^{h}+b\cdot(E-E_{0})+c\cdot(E-E_{0})^{2} & E\geq E_{0}\,.
  \end{cases}
  \label{eq:munewh}
 \end{eqnarray}
The temperature dependence of the individual parameters $par_i$ of Eqs.~\ref{eq:munewe},~\ref{eq:munewh} is expected to follow a power law:
\begin{eqnarray}
   par_{i}(T) & = & par_{i}(\unit[T=300]{K})\cdot \Big(\frac{\unit[T]{[K]}} {\unit[300]{K}}\Big)^{\alpha_{i}}\,.
  \label{eq:parT}
 \end{eqnarray}


\section{Measurements and analysis methods}\label{sec:meas}

 \paragraph{Sensors}

 The sensors investigated are three  pad diodes (p$^+$-n-n$^+$ and n$^+$-p-p$^+$ ) produced by two vendors on <100> silicon with bulk doping of 0.8 to 3.6$\cdot 10^{12}$~cm$^{-3}$.
 Their thicknesses are 200 and $\unit[287]{\upmu m}$, which were determined to an accuracy of 1~\%. Their areas are 4.4 and $\unit[25]{mm^2}$. A detailed description of the individual diodes is given in Ref.~\cite{KS2015}.

 \paragraph{Measurement set-ups}

 The transient current technique (TCT) was employed to measure the charge carrier drift velocities.
 Sub-nanosecond laser light pulses were used to generate electron-hole pairs in the sensors, which drift in the electric field to the electrodes.
 The current induced by the drifting charges is amplified and recorded by an oscilloscope with 5~Gsamples/s.
 Light with a wavelength of 675~nm, which has an absorption length of a few $\upmu$m in silicon, has been injected from both sides in order to produce signals dominated by either electrons or holes.
 Light with a wavelength of 1063~nm, which has an absorption length larger than the sensor thicknesses, produces signals with approximately equal contributions from electrons and holes.
 Most measurements were performed at the University of Hamburg, and  control measurements were also made at the CERN-SSD Lab~\cite{CERNSSDLAB}.

 \paragraph{Fit method}
 \label{sec:tof}

\begin{figure}[!ht]
  \centering
   \begin{subfigure}[a]{7.5cm}
    \includegraphics[width=7.5cm]{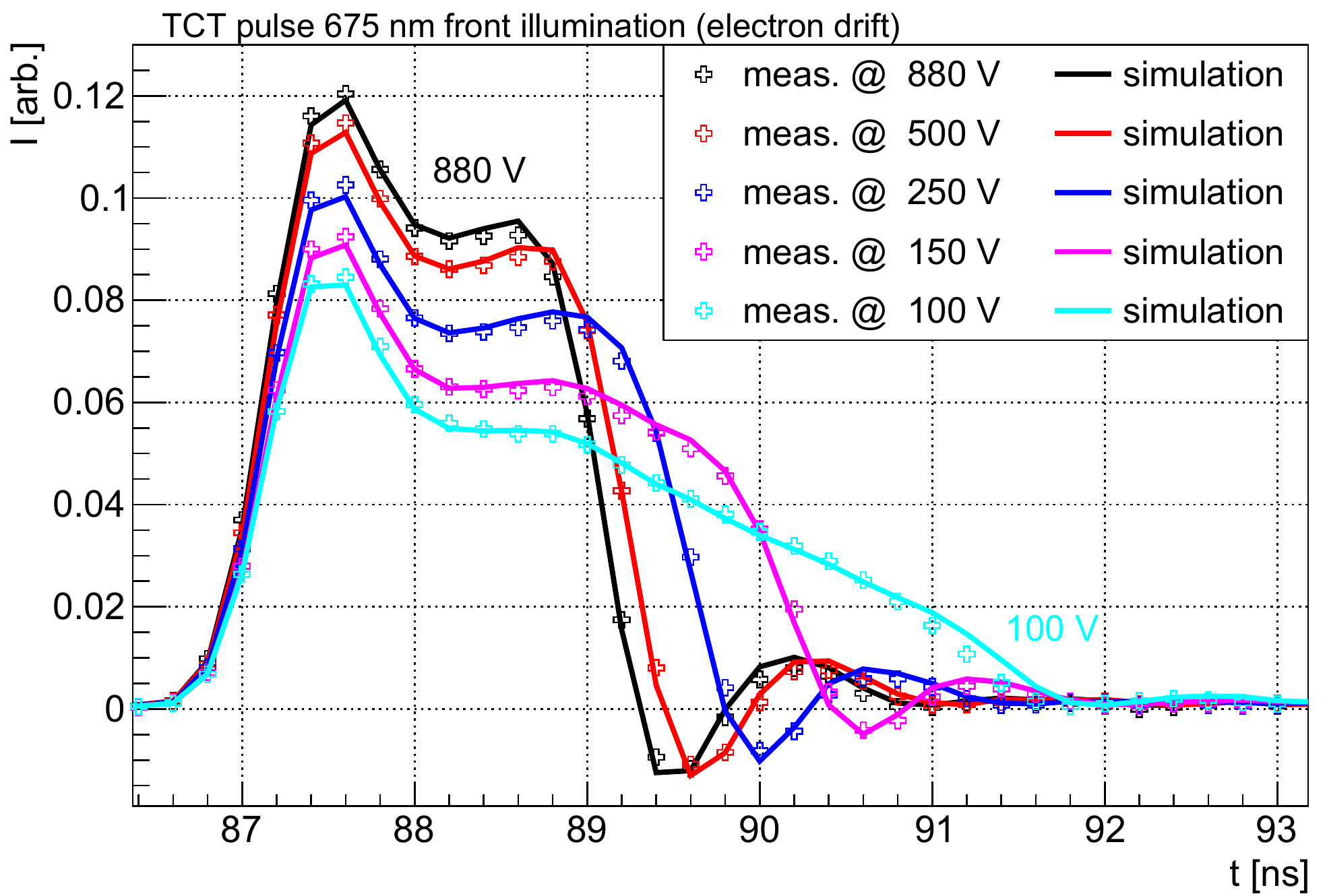}
    \caption{ }
     \label{fig:E233K}
   \end{subfigure}%
    ~
   \begin{subfigure}[a]{7.5cm}
    \includegraphics[width=7.5cm]{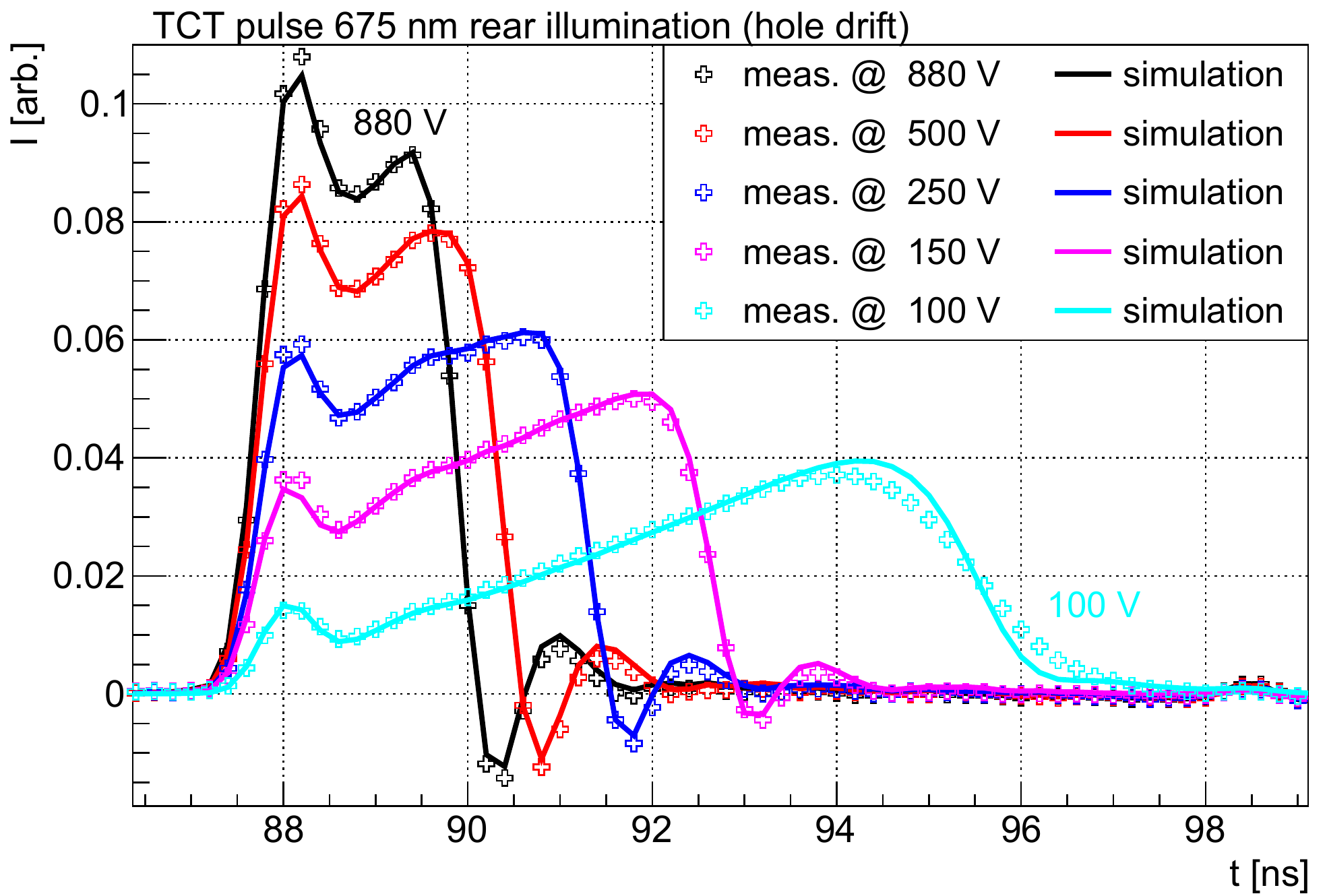}
    \caption{ }
     \label{fig:H233K}
   \end{subfigure}
    ~
   \begin{subfigure}[a]{7.5cm}
    \includegraphics[width=7.5cm]{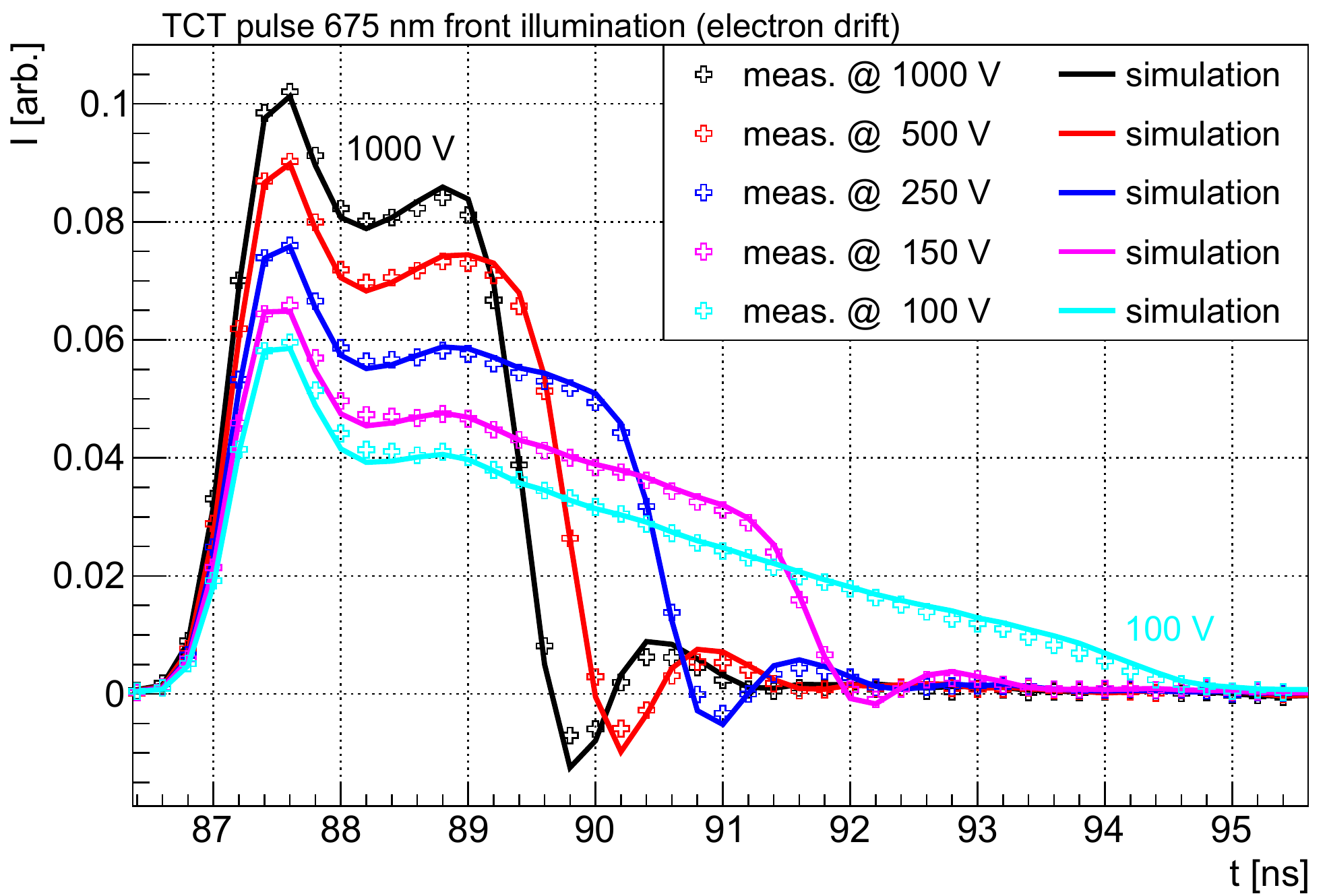}
    \caption{ }
     \label{fig:E333K}
   \end{subfigure}%
    ~
   \begin{subfigure}[a]{7.5cm}
    \includegraphics[width=7.5cm]{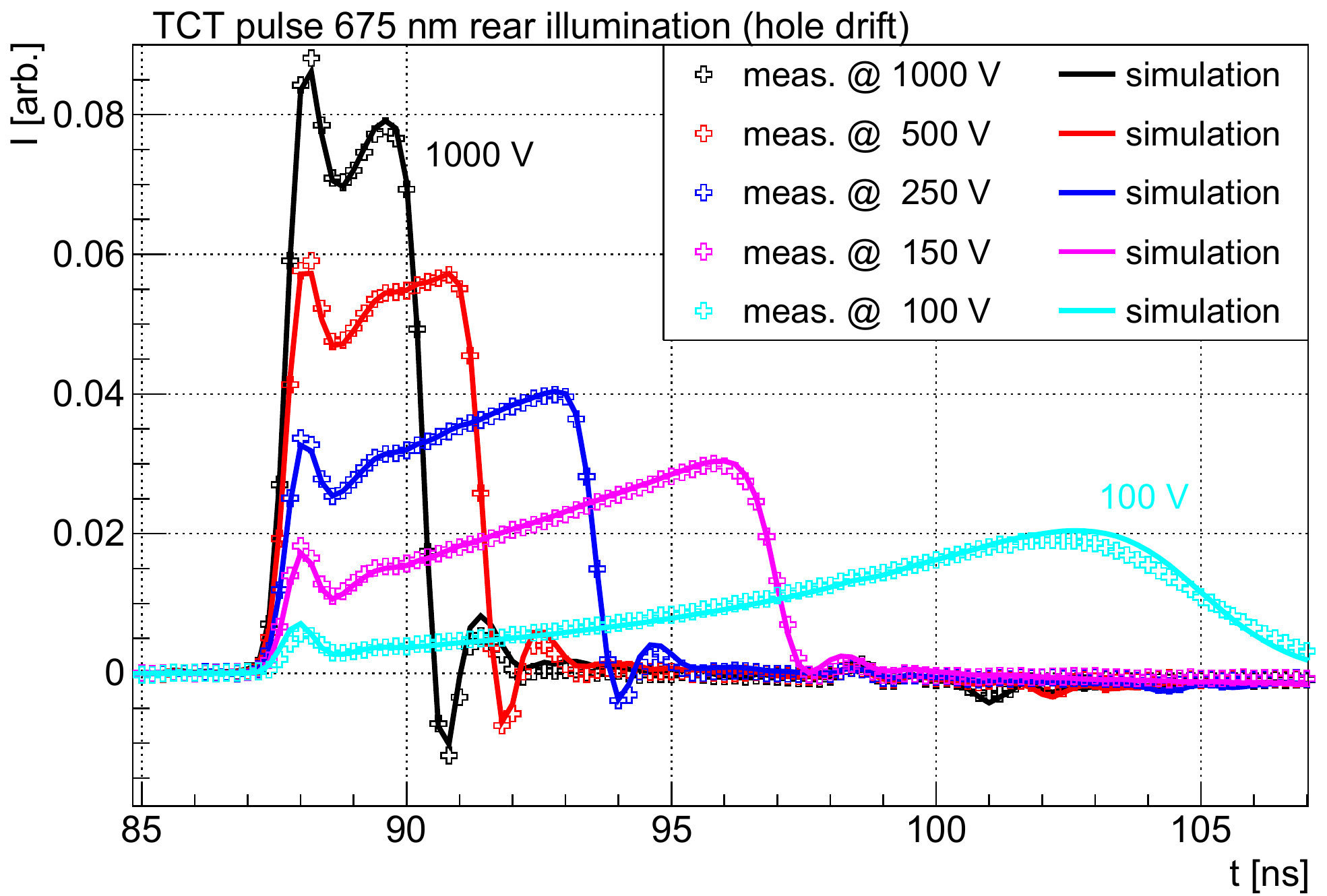}
    \caption{ }
     \label{fig:H333K}
   \end{subfigure}%
   \caption{\,Measured current transients (crosses) and the simulated ones (solid lines) using the parameters of Tab.~\protect\ref{tab:Tempdeppar1}(a) for a 200~$\upmu$m thick n-type diode using 675~nm laser light and bias voltages between 100 and $\unit[1000]{V}$.
    (a) Electron signal at $\unit[233]{K}$ (front-side illumination),
    (b) hole signal at $\unit[233]{K}$ (rear-side illumination),
    (c) electron signal at $\unit[333]{K}$, and
    (d) hole signal at $\unit[333]{K}$.}
  \label{fig:Global-fit}
 \end{figure}

 In the fit method, the current transients are simulated with the mobility parameterizations discussed in Sect.~\ref{sec:mobility}, convolved with the electronic transfer function and fitted to the measurements for determining the values of the mobility parameters.
 Electron-hole pairs are generated according to the temperature-dependent light attenuation on a 50~nm grid, the time step for the drift is 10~ps, and diffusion is taken into account using the Einstein relation.
A uniform doping is assumed resulting in an electric field with a constant gradient normal to the surface the pad diode.
 Figure~\ref{fig:Global-fit} shows the comparison of the fit results to the measured transients for a 200~$\upmu$m thick n-type diode using the 675~nm laser for  bias voltages between 100 and $\unit[1000]{V}$ and temperatures of 233 and 333~K.
 The agreement is excellent with the exception of the hole drift data at 100~V, which is just above the depletion voltage of this diode. Here plasma effects~\cite{Tove1967} become relevant due to the high carrier density in very low field.

\paragraph{Time-of-flight method}
\label{sec:tof}

 A time-of-flight (tof) method is used to cross check the fit method.
 The measured transients with the 675~nm laser light are interpolated and $t_{tof}$, the difference between the time of the maximum slope at the rise and at the fall of the pulse, is determined.
 For the mobility parametrization $1/\mu(E)\propto a_1+a_2\cdot E$, as in Eqs.~\ref{eq:newparlin},~\ref{eq:munewe}, the electric field strength corresponding to the drift velocity $v(E_{tof})=\frac{w}{t_{tof}}$ is $1/E_{tof} = \langle 1/E(x) \rangle$, where $w$ denotes the sensor thickness.
 For uniform bulk doping, where the electric field strength depends linearly on the distance to the electrodes, $E_{tof}=(E_{max}-E_{min})/\ln(E_{max}/E_{min})$, with $E_{max}$ the maximum and $E_{min}$ the minimum electric field strength in the sensor.

\section{Results}\label{sec:res}

 The tof and fit results agree within 1~\% for bias voltages  approximately 30~V above the depletion voltage.
 The results obtained from the different sensors and the different set-ups agree within 4~\%.
 The agreement between the results from the global fit for all temperatures simultaneously to the fits at the individual temperatures is within 2~\%.
 Table~\ref{tab:Tempdeppar1}(a) gives the parameters from the global fit.
 We estimate that for electric field strengths between 2.4 and 50~kV/cm the uncertainty of the drift velocities calculated using these parameters is 2.5~\% for electrons and 5~\% for holes.
 The comparison of the measured transients to the simulated ones using the parameters from the global fit for the lowest and highest temperatures for different bias voltages is shown in Figure~\ref{fig:Global-fit}.

 In order to provide a parametrization for field strengths between 0 and 50~kV/cm we performed a fit in which the low-field mobilities $\mu_{0,Jac}^{e,h}$ and their temperature dependencies $\alpha_i$ have been fixed to the values given in Ref.~\cite{jac1977}.
 The results are given in Table~\ref{tab:Tempdeppar1}(b).
 As there is some tension between the low-field mobilities from Ref.~\cite{jac1977} and our values, we recommend to use the values given in Table~\ref{tab:Tempdeppar1}(a) for $E\geq2.4$~kV/cm.

\begin{table}[tbp]
\begin{centering}
\begin{minipage}[t]{0.45\textwidth}%
\begin{center}
\begin{tabular}{|c||c|c|}
\hline
\multicolumn{3}{|c|}{$\unit[E =(2.4-50)]{kV/cm}$}\tabularnewline
\hline
$par_{i}$ & $par_{i}(\unit[T=300]{K})$  & $\alpha_{i}$\tabularnewline
\hline
$\mu_{0}^{e}$  & $\unit[1430]{cm^{2}/Vs}$  & $-1.99$\tabularnewline
\hline
$v_{sat}^{e}$  & $\unit[1.05\cdot10^{7}]{cm/s}$  & $-0.302$\tabularnewline
\hline
\hline
$\mu_{0}^{h}$  & $\unit[457]{cm^{2}/Vs}$  & $-2.80$\tabularnewline
\hline
$b$  & $\unit[9.57\cdot10^{-8}]{s/cm}$  & $-0.155$\tabularnewline
\hline
$c$  & $\unit[-3.24\cdot10^{-13}]{s/V}$  & $-$\tabularnewline
\hline
$E_{0}$  & $\unit[2970]{V/cm}$  & $5.63$\tabularnewline
\hline
\end{tabular}
\par\end{center}

\begin{center}
(a)
\par\end{center}%
\end{minipage}\hfill{}%
\begin{minipage}[t]{0.45\textwidth}%
\begin{center}
\begin{tabular}{|c|c|c|}
\hline
\multicolumn{3}{|c|}{$\unit[E =(0-50)]{kV/cm}$}\tabularnewline
\hline
$par_{i}$ & $par_{i}(\unit[T=300]{K})$  & $\alpha_{i}$\tabularnewline
\hline
$\mu_{0,Jac}^{e}$  & $\unit[1530]{cm^{2}/Vs}$  & $-2.42$\tabularnewline
\hline
$v_{sat}^{e}$  & $\unit[1.03\cdot10^{7}]{cm/s}$  & $-0.226$\tabularnewline
\hline
\hline
$\mu_{0,Jac}^{h}$  & $\unit[464]{cm^{2}/Vs}$  & $-2.20$\tabularnewline
\hline
$b$  & $\unit[9.57\cdot10^{-8}]{s/cm}$  & $-0.101$\tabularnewline
\hline
$c$  & $\unit[-3.31\cdot10^{-13}]{s/V}$  & -\tabularnewline
\hline
$E_{0}$  & $\unit[2640]{V/cm}$  & $0.526$\tabularnewline
\hline
\end{tabular}
\par\end{center}

\begin{center}
(b)
\par\end{center}%
\end{minipage}
\par\end{centering}

\caption{
 Mobility parameters for <100> silicon obtained from the fit to the data using Eq.~\protect\ref{eq:munewe} for the electron mobility, Eq.~\protect\ref{eq:munewh} for the hole mobility, and Eq.~\protect\ref{eq:parT} for the temperature dependence.
 The mean electric field strength in the sensors was $\unit[(2.4-50)]{kV/cm}$ and the temperature $\unit[(233-333)]{K}$.
 (a) Parameters from the global fit of the data.
 (b) Parameters from the global fit with $\mu_{0,Jac}^{e,h}$ and the corresponding $\alpha_i$ of Ref.~\protect\cite{jac1977} fixed. We recommend to use (a) for $E \geq 2.4$~kV/cm.\label{tab:Tempdeppar1}
}
\end{table}

\begin{figure}
  \centering
   \begin{subfigure}[a]{7.5cm}
    \includegraphics[width=7.5cm]{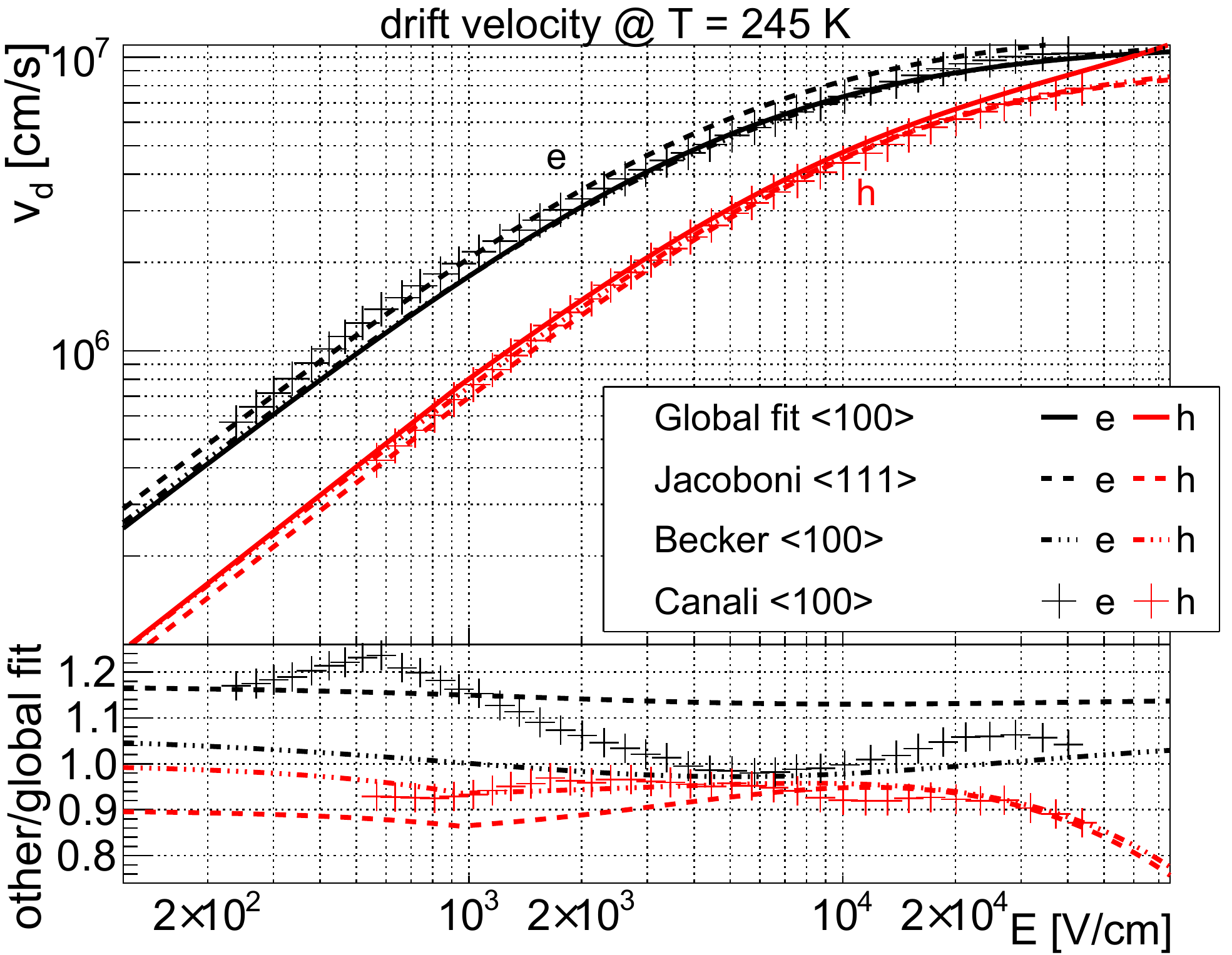}
    \caption{ }
     \label{fig:245K_vgllit}
   \end{subfigure}%
    ~
   \begin{subfigure}[a]{7.5cm}
    \includegraphics[width=7.5cm]{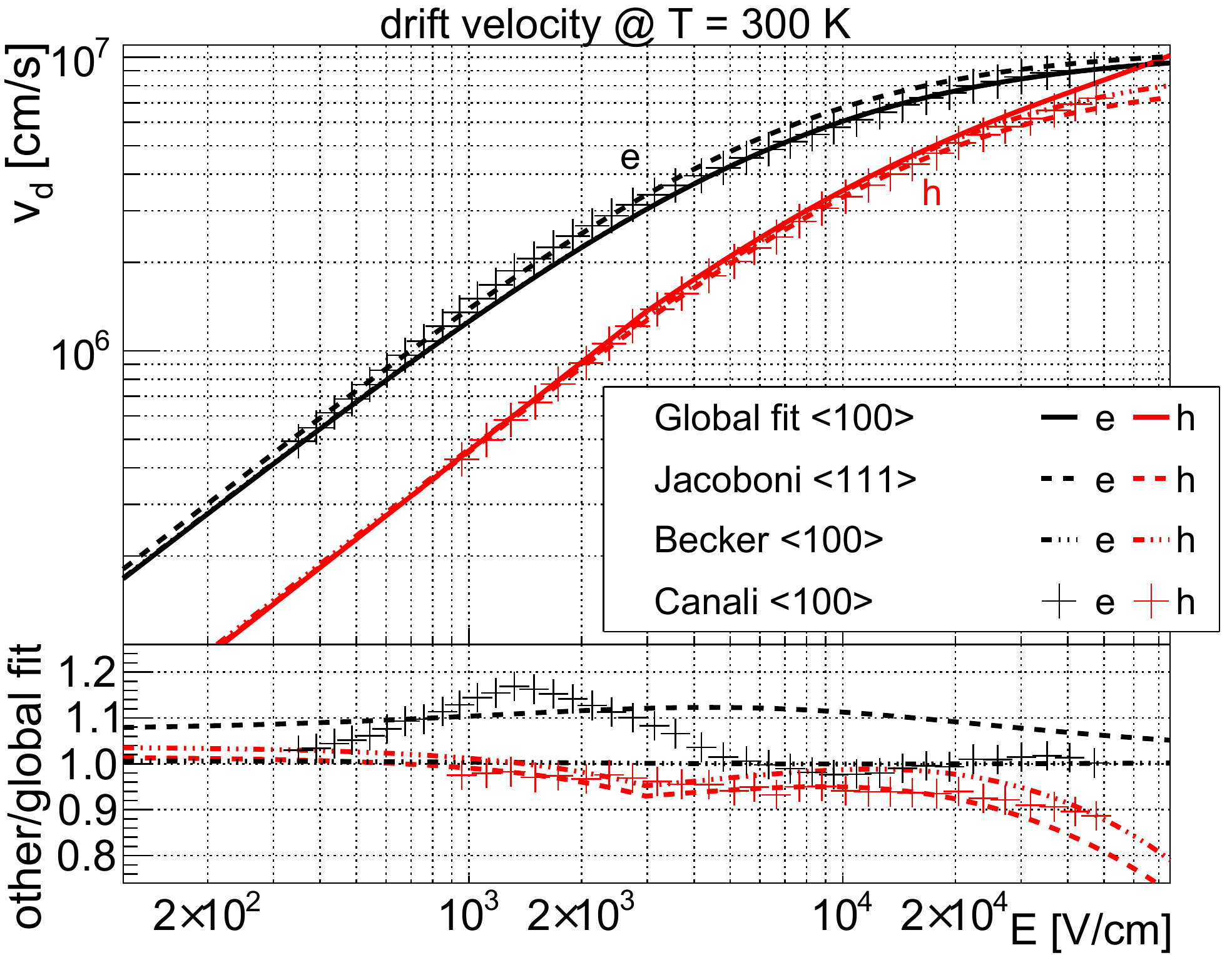}
    \caption{ }
     \label{fig:300K_vgllit}
    \end{subfigure}%
   \caption{
   Comparison of the drift velocities of electrons (e) and holes (h) from the global fit results of the data to the published data.
   The upper part shows the drift velocities and the lower part their ratios for
   (a) $\unit[245]{K}$, and
   (b) $\unit[300]{K}$.   }
  \label{fig:Comp-fit-lit}
 \end{figure}

 Figure~\ref{fig:Comp-fit-lit} compares the drift velocities calculated using the parameters of Table~\ref{tab:Tempdeppar1}(a) for the <100> direction to the values of Jacoboni et al.~\cite{jac1977} for  the <111>, of Becker et al.~\cite{bec2011} for the <100>, and the tof data of Canali et al.~\cite{canali1971drift} for the <100> direction.
 As the differences between our <100> results and the <111> results are typically 10~\% for electrons and up to 20~\% for holes at high electric field strengths, they should not be used for simulating <100> silicon.
 However, this is usually done, because for strong fields no parameterizations for the <100> direction are available.
 The <100> values of Becker et al. agree with our results within 6~\%, except for the hole velocity at strong fields, where these data had no sensitivity.
 The Canali tof values agree with our results to within better than 10~\% for $E \geq 2.4$~kV/cm except for the hole velocity at strong fields.

\section{Summary}\label{sec:sum}

 The drift velocities of electrons and holes for high ohmic silicon with <100> orientation have been measured for electric field strengths between 2.4 and 50~kV/cm and temperatures between 233 and 333~K.
 Compared to previous measurements, significant differences of up to 20~\% have been found.
 Parameterizations of the drift velocities with an estimated uncertainty of 2.5~\% for electrons and 5~\% for holes in the field strength and temperature range of the measurements are provided.
 In addition, a less precise parametrization for electric field strengths between 0 and 50~kV/cm is presented.

\acknowledgments

 The authors would like to thank J. Becker for making available his current transient simulation program, and E. Fretwurst, E. Garutti and J. Schwandt for stimulating discussions.
 We also thank M. Moll for giving us access to the CERN-SSD TCT set-up, and  H. Neugebauer and Ch. Gallrapp, who helped in preparing and performing the measurements at CERN.
 The authors also thank the HGF Alliance Physics at the Terascale for funding the Hamburg TCT set-up.

\bibliographystyle{unsrtnat}

\begin{thebibliography}{99}

\bibitem{jac1977}
C.~Jacoboni, C.~Canali, G.~Ottaviani, and A.~Alberigi~Quaranta.
 A review of some charge transport properties of silicon.
 \emph{Solid-State Electronics}, 20(2):77--89, 1977.


\bibitem{bec2011}
J.~Becker, E.~Fretwurst, and R.~Klanner.
 Measurements of charge carrier mobilities and drift velocity
  saturation in bulk silicon of <111> and <100> crystal orientation at high
  electric fields.
 \emph{Solid-State Electronics}, 56(1):104--110, 2011.


 \bibitem{KS2015}
R.~Klanner and C.~Scharf.
 Measurement of the drift velocities of electrons and holes in high-ohmic <100> silicon.
 \emph{Nuclear Instruments and Methods in Physics Research Section A:
  Accelerators, Spectrometers, Detectors and Associated Equipment}, 799:81--89, 2015.


 \bibitem{scharf2014thesis}
C.~Scharf.
\newblock  Measurement of the drift velocities of electrons and holes in
  high-ohmic <100> silicon.
\newblock {Master thesis}, University of Hamburg, 2014.
\newblock {DESY-THESIS-2014-015}.


\bibitem{transferf}
C.~Scharf and R.~Klanner.
\newblock Determination of the electronics transfer function for current
  transient measurements.
\newblock {\em Nuclear Instruments and Methods in Physics Research Section A:
  Accelerators, Spectrometers, Detectors and Associated Equipment}, 779:1--5, 2015.


\bibitem{caughey1967carrier}
R.~E. Thomas.
\newblock Carrier mobilities in silicon empirically related to doping and
  field.
\newblock {\em Proceedings of the IEEE}, 55(12):2192--2193, Dec 1967.


\bibitem{canali1971drift}
C.~Canali, G.~Ottaviani, and A.~Alberigi~Quaranta.
\newblock Drift velocity of electrons and holes and associated anisotropic
  effects in silicon.
\newblock {\em Journal of Physics and Chemistry of Solids}, 32(8):1707--1720,
  1971.


\bibitem{trofimenkoff1965field}
F.~N. Trofimenkoff.
\newblock Field-dependent mobility analysis of the field-effect transistor.
\newblock {\em Proceedings of the IEEE}, 53(11):1765--1766, 1965.


  \bibitem{CERNSSDLAB}
{CERN SSD (Solid State Detector) lab of the CERN Physics Department}, 2015.
\newblock \href{http://www.cern.ch/ssd/}{http://www.cern.ch/ssd/}.


\bibitem{Tove1967}
 P.~A.Tove and W.~Seibt,
 \newblock Plasma effects in semiconductor detectors.
 \newblock {\em Nuclear Instruments and Methods in Physics Research Section A:
  Accelerators, Spectrometers, Detectors and Associated Equipment}, 51(2):261 --  269, 1967.


\end{thebibliography}

\end{document}